\documentclass[aps,preprint,groupedaddress,showpacs,floatfix]{revtex4}

\usepackage{epsfig}
\usepackage{amsmath}
\usepackage{amssymb}
\newcommand{\ua}{\uparrow}
\newcommand{\da}{\downarrow}

\begin{document}

\title{Zeno Quantum Gates in Semiconductor Quantum Dots}

\author{K. J. Xu, Y. P. Huang, M. G. Moore, and C. Piermarocchi}
\address{Department of Physics \& Astronomy, Michigan State
University, East Lansing, 48824}
\date{\today}

\begin{abstract}
We propose a scheme for a two-qubit conditional phase gate by
quantum Zeno effect with semiconductor quantum dots. The system
consists of two charged dots and one ancillary dot that can perform
Rabi oscillations under a resonant laser pulse. The quantum Zeno
effect is induced by phonon-assisted exciton relaxation between the
ancillary dot and the charged dots, which is equivalent to a
continuous measurement. We solve analytically the
master equation and simulate the dynamics of the system using a
realistic set of parameters. In contrast to standard schemes, larger
phonon relaxation rates increase the fidelity of the operations.
\end{abstract}

\maketitle

The Quantum Zeno effect (QZE)~\cite{Misra77,Chiu77,Peres80,Itano90}
occurs when a rapid sequence of measurements is performed on a
slowly evolving quantum system, with the result that the system is
{\it frozen} in its initial state. An analogous effect occurs when a system is strongly coupled to a reservoir, as the transfer of information from the system into the reservoir mimics a continuous measurement. If coupling to the reservoir can be made contingent on the joint quantum state of two qubits, the QZE can be used in conjunction with control pulses to efficiently drive the qubits into an entangled state. This approach have been discussed
within the framework of interaction free
measurements~\cite{Kwiat95,Horodecki01,Huang08}, decoherence free
subspaces~\cite{Rasetti00,Facchi02}, as well as counterfactual quantum
computation~\cite{Hosten06}. Proposed physical implementations vary from
purely photonic systems~\cite{Franson04,Leung06,Myers07},  to atom-cavity systems \cite{Beige00,Huang08}, and
superconducting qubits~\cite{nori08}.

Following a generalized QZE phase-gate described in~\cite{Huang08}, we have devised
a two-qubit conditional phase gate using electron spins in
semiconductor quantum dots. This system has the advantage that decoherence rates are
of the order of picoseconds, which can lead to significant improvements in gate time and/or fidelity relative to atomic systems.
We consider a system composed of three
quantum dots (QDs), two of which are singly charged with electrons. 
The spins of these two electrons are then the logical qubits on which the phase-gate acts.
A laser field is then applied, tuned to the exciton resonance of the uncharged dot.
The energy levels and laser polarization are chosen so
that an exciton in the neutral dot can relax to the neighboring dots by a spin-conserving dissipative
phonon-assisted process in which the exciton is annihilated into a virtual photon and recreated
as an exciton and a phonon in a neighboring dot~\cite{Rozbicki08}. 
The emission of a phonon clearly carries away the information that at least one qubit spin did not match the electron spin of the exciton, otherwise the decay mechanism would have been Pauli blocked. Thus the possibility of phonon emission is equivalent to a continuous partial
measurement~\cite{Hohenester07,Huang08} of the collective spin state of the two qubits. 

Despite the widely-held belief
that decoherence must always be minimized in quantum information
processing, it has been known for some time~\cite{Beige00} that
decoherence can in principle be harnessed to generate high-fidelity
entanglement by use of the QZE. In our scheme, the QZE effect occurs when the strong dissipation rate of the exciton state suppresses the laser induced Rabi oscillations in the neutral dot, effectively freezing it in its ground state. As the dissipation mechanism is subject to Pauli blocking, the
spin-qubits in the charged electrons can thus be seen as {\it quantum
switches}. If both switchs are set to {\it off} (all electron spins aligned up) then the measuring device (the environment)
is inactivated and the QZE does not occur. In this way, the driving laser can create a tri-partite entangled state involving the two spin qubits and the state of the neutral dot.

In order to predict the fidelity of state formation, we will
use parameters appropriate for vertically grown (In,As)Ga/GaAs self-assembled
quantum dots. Structures with vertically coupled neutral and charged
quantum dots have been recently demonstrated~\cite{Robledo08}. Henceforth, 
the central neutral dot will be QD1, and  the two lateral
charged dots will be referred to as QD2 and QD3. Absorption of a photon creates  an exciton state in QD1, and
trion states in QD2 and QD3. We assume that the ground trion energies
in the two lateral dots are similar, and are lower than the ground
exciton energy in the central dot, so that phonon-mediated exciton relaxation is energetically allowed. 
In the absence of the QZE, the driving laser will induce Rabi
Oscillations between the zero- and one-exciton states of QD1.  Assuming that the
driving laser field is $\sigma_{-}$ polarized, the standard selection rules lead to the creation of
an exciton with electron spin
up($+\frac{1}{2}$) and a heavy hole spin down($-\frac{3}{2}$) in QD1. Due to the difference
between the exciton and trion energy, QD2 and QD3 are far-detuned and thus not driven by the laser.

\begin{figure}
       \begin{center}
       \includegraphics[width=0.8\textwidth]{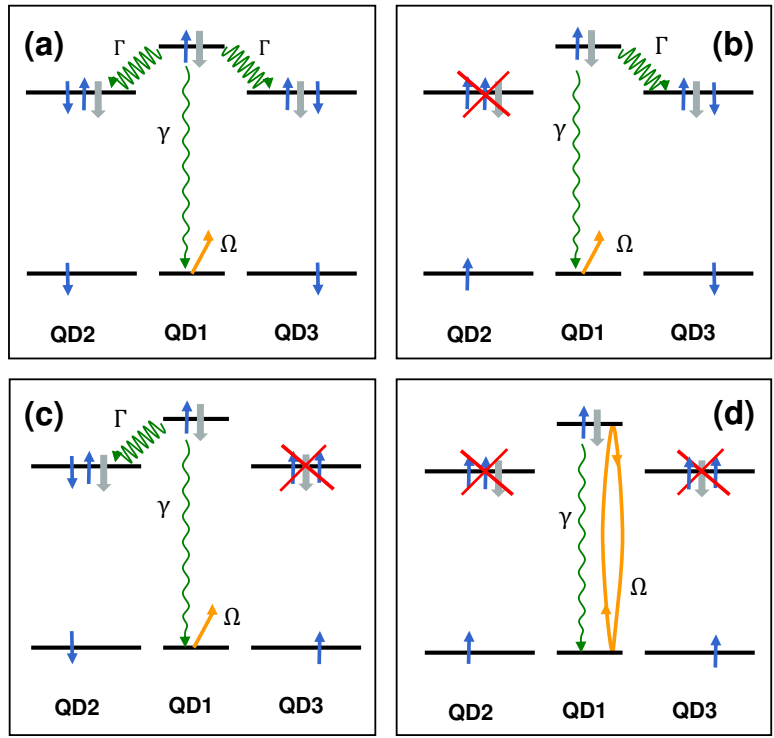}
       \end{center}
       \caption{(Color online) Scheme of the dynamics of the system under different
       initial states. The narrow blue (bold gray) arrow represents the electron
       (hole) spins. The energy levels for QD1 are the empty dot
       (lower) and the first exciton level (upper), while the energy levels
       in QD2 and QD3 are charged dot ground states (lower) and trion
       states (upper). Figures (a)-(d) correspond to the four possible initial states of the two qubits. 
      (a)-(c) If either electron in QD2 and QD3 is spin down ,
       the exciton in QD1 can decay into the neighboring dots, in which case
       the QZE prevents the Rabi Oscillation in QD1.  (d) If both
       decay channels are closed, QD1 will undergo a $2\pi$-Rabi oscillation. The photon emission rate $\gamma$ is too weak to induce a QZE.}
       \label{sys}
\end{figure}

There are several exciton decay mechanisms that can spoil the Rabi oscillations. Many such
processes depend on the intensity of the laser and have been
experimentally characterized~\cite{stievater01,wang05}. In the weak
excitation limit, phonon-mediated processes are dominant.
The role of the phonon is to carry away excess energy.
As phonon-assisted relaxation of a {\it single} carrier between two dots via tunneling
is exponentially suppressed for QD separations of several nm,
we focus on the aforementioned {\it exciton} relaxation via long-range phonon-assisted
excitation transfer~\cite{Rozbicki08}.
For a weak resonant $2\pi$ laser pulse with Rabi energy $\Omega
\ll \Gamma$, where $\Gamma$ is the phonon emission rate, then
a single spin-down electron will be sufficient to impose the QZE and
freeze the system. When both spins are up, the
$2\pi$ pulse will give an overall $\pi$ phase to the wave-function
of the system. A schematic view of the different possible QZE scenarios for different initial states is shown in Fig.\ref{sys}.
In any case, the final state of the ancillary system will be $|0\rangle_a$, so that the $\pi$ phase shift is effectively imprinted
only on the $|\uparrow\uparrow\rangle_{23}$ state, thus realizing a two-qubit phase gate.

For concreteness, we
will assume that the phonon-assisted excitation transfer is the
dominant dissipation channel from QD1 to QD2 or QD3. However, the
scheme for the gate and our analytical results, can be easily
adapted to the case in which QDs are close and the phonon emission
involves only the electron. In this short-range case the energy
levels of the central and lateral dots have to be engineered so that hole transfer is forbidden.

The quantum state of the system can be expressed in the
triple-particle basis $|\lambda\sigma\sigma'\rangle=|\lambda\rangle_a\otimes|\sigma
\sigma^\prime\rangle_{23}$. Here, $|\lambda\rangle_a$
($\lambda=0,1,2,3$) represents the state of the ancillary electron-hole pair created in QD1,
where $\lambda=0$ is the vacuum state with no exciton, while $\lambda=1,2,3$ indicates
the exciton residing in QD1,QD2, or QD3, respectively. $|\sigma\sigma'\rangle_{23}$ represents the combined state of the
two logical qubits, with $\sigma,\sigma^\prime=\uparrow,\downarrow$
indicating the spin (up or down) states of the electrons in QD2 and
QD3, respectively. We note that during
the Zeno gate operation, the probability to excite more than one
exciton is exponentially small, and hence only the single-excitation
case is considered in the present discussion.
The states
$|2\!\uparrow\! \sigma'\rangle$ and
$|3\,\sigma\!\uparrow\rangle$ are
forbidden by the Pauli exclusion principle, and therefore excluded from our model.

Our goal is to
realize a two-qubit phase gate for the electron spins in QD2 and QD3, with the electron-hole pair acting only as
an ancillary system.
Ideally, such a gate transforms
an initial logical state $|\Psi_{i}\rangle_{23}$ of QD2 and QD3 to the
final state $\hat{U}_\pi |\Psi_{i}\rangle_{23}$, with the $\pi$-phase gate operator defined via
$\hat{U}_\pi |\sigma\sigma'\rangle_{23}
=(1-2\delta_{\sigma,\uparrow} \delta_{\sigma',\uparrow})|\sigma
\sigma'\rangle_{23}$. The ancillary system, initially prepared in
$|0\rangle_a$ state, becomes entangled with the logical qubits
during the $2\pi$ pulse, becoming once again disentangled by the end of the pulse.
In practice, the
ancillary qubit could still be entangled with the logical
qubits after the gate operation, so that the final density matrix representing
QD2 and QD3 is obtained by tracing over the $|\lambda\rangle_a$ states.

The system's Hamiltonian is given by
\begin{equation}
H=\epsilon _{2}c_{2}^{\dag }c_{2}+\epsilon _{3}c_{3}^{\dag
}c_{3}+\frac{\Omega}{2} (c_{1}+c_{1}^{\dag}) \label{maineqn}
\end{equation}
Here, the $c_{i}^{\dag }(c_{i})$ is the exciton creation
(annihilation) operator, with $i=1,2,3$ labeling the three QDs.
$\Omega$ is the Rabi strength of the driving laser. In contrast to the state-selectivity of the phonon-mediated relaxation process, the decay of the exciton in QD1 via
spontaneous photon emission is independent of QD2 and QD3 states, and will only cause
the exciton to relax back to the initial $|0\rangle_a$ state. This is the primary source of error in the gate operation, and is mitigated by choosing $\Omega\gg \gamma$, where $\gamma$ is the exciton spontaneous photon-emission rate.

To model the system's dynamics, we employ the standard Linblad formalism \cite{Mey01} to arrive at the master equation
\begin{equation}
\label{meq}
 i\frac{\partial \rho }{\partial t}=-[\rho
,H]+i\mathcal{L} [ \rho ],
\end{equation}
where $\rho$ is the density operator for the system. The superoperator $\mathcal{L}$ is given by
\begin{equation}
\mathcal{L} [ \rho ]=\frac{1}{2}[L_{\gamma}\rho L_{\gamma}^{\dag
}+L_{\Gamma}\rho L_{\Gamma}^{\dag }-(L_{\gamma}^{\dag }L_{\gamma}+
L_{\Gamma}^{\dag }L_{\Gamma})\rho +H.c.],
\end{equation}
where $L_{\gamma }=\sqrt{\gamma }c_{1}$ describes spontaneous
photon decay in QD1, and $L_{\Gamma }=\sqrt{\Gamma }(c_{2}^{\dag }c_{1}+c_{3}^{\dag }c_{1})$ describes
phonon-assisted dissipation from QD1 to QD2 and QD3.
We note that aside from the photon- and phonon- relaxation
channels, other channels could as well be characterized
by generalizing the $\Gamma$-terms to include any spin-selective relaxation channels,
while $\gamma$-terms to include spin-independent ones.

During the gate operation, the system is initially in the state
$\rho_{i}=|\Psi_i\rangle\langle \Psi_i|\otimes |0\rangle_a\langle
0|_a$, and then evolves under equation (\ref{meq}) for a duration of
$t=2\pi/\Omega$, resulting in a final density $\rho_f$. The fidelity is
defined as
\begin{equation}
\label{fidelity}
    \mathcal{F}=tr\{\rho_f \hat{U}_\pi |\Psi_i \rangle_{23} \langle \Psi_i|_{23} \hat{U}^\dag_\pi
   \otimes \hat{P}_1\},
\end{equation}
where $\hat{P}_1=|0\rangle_a\langle 0|_a+|1\rangle_a\langle 1|_a$.
This gives the probability that two logical qubits are in the proper phase-gate output state with the electron-hole pair remaining in QD1. This later condition is required because relaxation of the exciton to either QD2 or QD3 results in a doubly-charged dot, thus spoiling the separability of the two-qubit subsystem.

Before presenting numerical results, we first seek approximate analytical
solutions to the dissipative dynamics of equation (\ref{meq}).
Defining density matrices $\rho_{mn}=\langle m |_{23} \hat{P}_1 \rho
\hat{P}_1 |n \rangle_{23}$, with $m,n=\uparrow\uparrow,\uparrow\downarrow,\downarrow\uparrow,\downarrow\downarrow$,
the master equation (\ref{meq}) can be divided into a set of
uncoupled equations, leading to
\begin{eqnarray}
\label{EqRd}
    \frac{\partial \rho_{mn}}{\partial t} &=&
    i[\rho_{mn},H_{0}] +\frac{\gamma}{2} \left(c_{1} \rho_{mn} c_{1}^{\dag}-
    c_{1}^{\dag}c_{1} \rho_{mn} + H.c.\right)  \nonumber \\
    & &-\alpha_{m} \Gamma
    c_{1}^{\dag}c_{1}\rho_{mn}-\alpha_{n} \Gamma \rho_{mn} c_{1}^{\dag}c_{1},
\end{eqnarray}
where $\alpha_{m}$ is a logical-qubit dependent parameter, defined
as $\alpha_{m}=0, \frac{1}{2}, \frac{1}{2}, 1$, for
$m=\uparrow\uparrow,\uparrow\downarrow,\downarrow\uparrow,\downarrow\downarrow$,
respectively.

Successful operation requires $\Gamma\gg\Omega$ to impose the QZE, while $\gamma\ll\Omega$ is required to suppress spontaneous photon-emission, the primary failure mechanism. Hence, the operational range of the present Zeno
phase gate is $\gamma\ll \Omega\ll \Gamma$. This separation of time-scales
enables us to solve Eq. (\ref{EqRd}) perturbatively. With the definition  $\rho^\lambda_{mn}=\langle \lambda m|\rho |\lambda n\rangle$,
the matrix elements of the final density are given to second order in $\frac{\gamma}{\Omega}$ and $\frac{\Omega}{\Gamma}$ by
$\rho^\lambda_{mn}=\mu^\lambda_{mn}\langle m|\Psi_i\rangle\langle\Psi_i|n\rangle$. The output
coefficients $\mu^\lambda_{mn}$ are given by Table 1,
with $f(x)=1-\frac{\pi}{2}x+\frac{3\pi^2}{50} x^2$,
$g(x)=\frac{\pi}{100}x+\frac{3\pi^2}{500}x^2$.
Note that the population and coherence dynamics  in the subspace $\lambda=2,3$ are completely decoupled from
the $\lambda=0,1$ subspace.
In fact, we only need
equations for the diagonal matrix elements with respect to the $\lambda=0,1$ subspace, as only they contribute to the fidelity (\ref{fidelity}).
We see from Table 1 that to leading order,
the gate output coefficients are consistent with
only the state $|\!\!\uparrow\uparrow\rangle_{23}$ having acquired a
$\pi$-phase shift, as desired for the
$\pi$-phase gate.

\begin{table}[htdp]
\caption{Output coefficients}
\label{table1}
\begin{center}
\begin{tabular}{c|c|c|c}
$m$ & $n$ & $\mu^0_{mn}$ & $\mu^1_{mn}$\\
\hline $\ua\ua$ & $\ua\ua$ & $(1+\frac{3\pi}{4}\frac{\gamma}{\Omega})^{-1}$ &
$1-(1+\frac{3\pi}{4}\frac{\gamma}{\Omega})^{-1}$ \\
$\ua\ua$ &  $\neq \ua\ua$ & $-f\!\!\left( \frac{\Omega}{\alpha_n\Gamma}\right)
\exp\left(-\frac{\pi}{2}\frac{\gamma}{\Omega}\right)$
 & $g\!\!\left(
\frac{\Omega}{\alpha_n\Gamma}\right) \left(\frac{\gamma}{\Omega}+\pi\frac{\gamma^2}{\Omega^2}\right)$ \\
$ \neq \ua\ua$ & $\ua\ua$ & $-f\!\!\left( \frac{\Omega}{\alpha_m\Gamma}\right)
\exp\left(-\frac{\pi}{2}\frac{\gamma}{\Omega}\right)$ &
 $g\!\!\left( \frac{\Omega}{\alpha_m\Gamma}\right)
 \left(\frac{\gamma}{\Omega}+\pi\frac{\gamma^2}{\Omega^2}\right)$ \\
$ \neq \ua\ua$ & $ \neq \ua\ua$ &
$\exp\left(-\frac{\pi}{2}\frac{\alpha_m+\alpha_n}{\alpha_m\alpha_n}
 \frac{\Omega}{\Gamma}\right)$ & $0$
\end{tabular}
\end{center}
\label{default}
\end{table}%

The fidelity defined in Eq. (\ref{fidelity}) is now explicitly given by
\begin{equation}
 \mathcal{F}=\sum_{ mn}
    (-1)^{\delta_{m,\uparrow\uparrow}+\delta_{n,\uparrow\uparrow}}
    (\mu^0_{mn}+\mu^1_{mn}) |\langle
    n|\Psi_i\rangle|^2 |\langle m| \Psi_i\rangle|^2,
\label{fm}
\end{equation}
which is dependent on $\gamma,\Omega,\Gamma$, as well as the initial
logical state $|\Psi_i\rangle$. In practice, while $\Gamma$ and
$\gamma$ are known parameters for a given QD system,
$|\Psi_i\rangle$ is in general arbitrary, making it impossible to
globally optimize $\Omega$ for all initial states. Nonetheless, for a given $\Omega$ we can always establish a lower bound for the fidelity.
Optimizing $\Omega$ to maximize the lower limit on the fidelity gives $\Omega_{opt}=\sqrt{\gamma\Gamma/8}$, with the optimized fidelity
obeying
\begin{equation}
\label{lowerbound}
    \mathcal{F}\ge \exp\left[-\frac{10}{3}\sqrt{\frac{\gamma}{\Gamma}}\right],
\end{equation}
which shows that the way to improve the gate fidelity is to decrease the ratio $\gamma/\Gamma$.

Considering recent theoretical calculations, the phonon-assisted
transfer rate between two QDs can be as fast as several tens of
picoseconds for favorable alignments \cite{Rozbicki08}. The lifetime
$\tau$ of the exciton in (In,As)Ga/GaAs QD is of the order of $1$ ns
\cite{Narvaez06}, which only marginally meets our operational
criteria. Nonetheless, $\tau$ can be significantly extended by
embedding the QD system into an optical cavity. In fact, $\tau\sim
10$ ns has been demonstrated in a recent experiment
\cite{Hennessy07}. For accessible parameters of $\Gamma=20$
ns$^{-1}$, $\gamma=0.08$ ns$^{-1}$, we find
$\Omega_{opt}=0.45$ ns$^{-1}$ and the lower bound for the fidelity
is $0.810$. For these parameters, the average fidelity \cite{Poyatos97, Piermarocchi02} is $\mathcal{F}_{avg}=0.85$. As we will descibe, a much higher fidelity can be obtained probabilistically by measuring the final state of the ancillary system to {\it herald} successful gate operation.

\begin{figure}
       \begin{center}
       \includegraphics[width=0.8\textwidth]{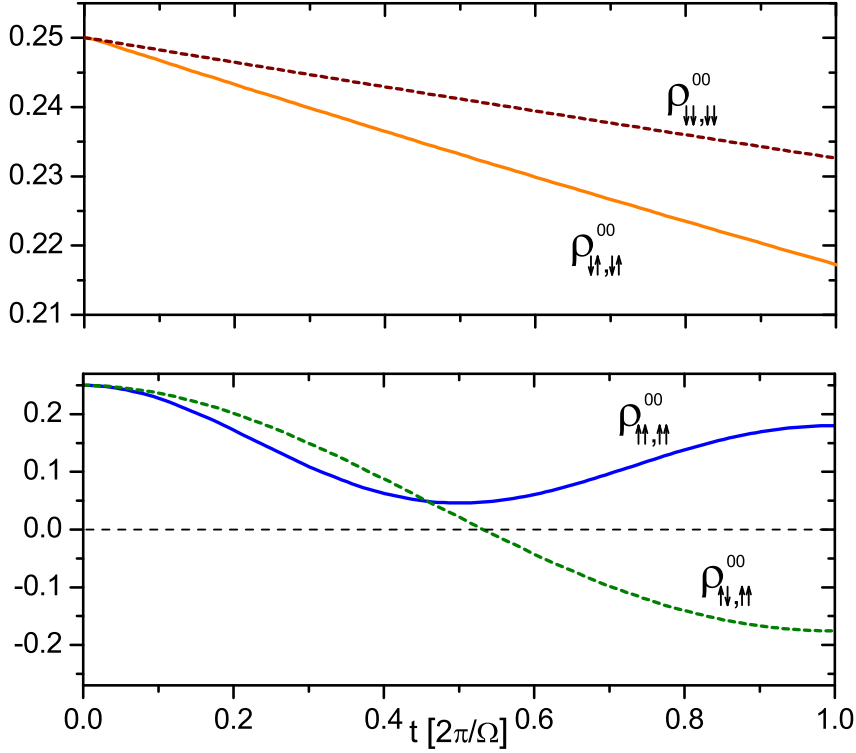}
       \end{center}
       \caption{Dynamical evolution of several density matrix
       elements for the initial state $|\Psi^0_i\rangle$
       during the gate operation via numerical simulation. }
       \label{result}
\end{figure}

To verify the analytical results, we now solve exactly the
dissipative dynamics Eq. (\ref{maineqn}) via numerical simulations.
For comparison, we choose $\Gamma=20$ ns$^{-1}$, $\gamma=0.08$
ns$^{-1}$, $\Omega=0.45$ ns$^{-1}$ and initial state
$|\Psi^0_i\rangle=\frac{1}{2}\left(|\ua\ua\rangle+|\ua\da\rangle+|\da\ua\rangle+|\da\da\rangle\right)$.
The dynamics of matrix elements
$\rho^{00}_{\uparrow\uparrow,\uparrow\uparrow},
\rho^{00}_{\uparrow\downarrow,\uparrow\uparrow}$ and
$\rho^{00}_{\downarrow\uparrow,\downarrow\uparrow},
\rho^{00}_{\downarrow\downarrow,\downarrow\downarrow}$ are shown in
Fig.\ref{result}. From the figure we can see that both
$\rho^{00}_{\uparrow\uparrow,\uparrow\uparrow}$ and
$\rho^{00}_{\uparrow\downarrow,\uparrow\uparrow}$ undergo damped
oscillation, due to $\gamma \ll \Omega$. At the end of the
$2\pi$-pulse, we find
$\rho^{00}_{\uparrow\uparrow,\uparrow\uparrow}=0.180$, and
$\rho^{00}_{\uparrow\downarrow,\uparrow\uparrow}=-0.176$, compared
with $0.176$ and $-0.175$ from the analytical results. In contrast,
both $\rho^{00}_{\downarrow\uparrow,\downarrow\uparrow}$ and
$\rho^{00}_{\downarrow\downarrow,\downarrow\downarrow}$ are shown to
be frozen in its initial state, due to QZE since $\Omega \ll
\Gamma$. The final values of
$\rho^{00}_{\downarrow\uparrow,\downarrow\uparrow}$ and
$\rho^{00}_{\downarrow\downarrow,\downarrow\downarrow}$ are found to
be $0.217$ and $0.233$, in agreement with the analytical results.
The fidelity from the numerical simulation is $\mathcal{F}_e=0.829$,
which is very close to the predicted value $0.831$ from equation
(\ref{fm}).

The gate fidelity can be further improved by  measuring the final state of the
ancillary electron-hole pair, which can `herald' successful gate operation. 
If it is detected in $|2\rangle_a,
|3\rangle_a$ or $|1\rangle_a$ states, which correspond to a trion
in QD2, QD3, or an exciton in QD1, failure is indicated. Only if the state $|0\rangle_a$ is obtained, is successful operation a possibility. 
Its fidelity, in this case is then
$\mathcal{F}_{h}=\mathcal{F}/(1-P_{f})$, where $P_{f}$ is the
failure probability. For the input state $|\Psi^0_i\rangle$, this
heralded fidelity is $\mathcal{F}_{h}=0.986$, a significant
improvement from the unheralded value of $0.829$. Similar
improvements are found for other input states. The reason for the large improvement is that the dominant failure mechanism is photon emission via exciton decay in QD1. This is most likely to occur at the halfway point of gate operation. This results in QD1 returning to $|0\rangle_a$ and begin a new Rabi oscillation cycle. In this scenario, only half of a Rabi cycle will have occurred, leaving QD1 in the exciton states. Thus failure due to photon emission will correlate highly with the ancillary system being found in state $|1\rangle_a$ at the end of gate operation.

In practice, one feasible way to measure the final state of the ancillary
electron-hole pair  is to apply two driven lasers
to the three QDs and detect the outcoming fluorescence photons. One
of the lasers is tuned to be resonant with the trion and charged
biexciton transition in QD2 and QD3, yet far detuned from other
transitions. Similarly, the other laser is resonant with the exciton
and biexciton transition in QD1, and as well far detuned from other
transitions. A trion in QD2 and QD3, or an exciton in QD1, will then
lead to resonance fluorescence, indicating the failure of the gate
operation. On the other hand, the absence of fluorescence photons
heralds the electron-hole being in the $|0\rangle_a$ state. We note
that in the non-fluorescence case, the logical qubits are preserved,
since they are only far off-resonantly coupled by the driven fields.

The results of this work demonstrate the possibility to realize a two-qubit
controlled phase gate via the QZE in (In,As)Ga/GaAs self-assembled
quantum dots.  Using experimental values for all parameters, the obtained fidelity
around $0.85$. If the final state of the exciton can successfully
be measured, the heralded fidelity
can be as high as $0.99$. The fidelity can be improved further only if the phonon-assisted exciton
transfer rate can and/or the lifetime
of the exciton in the ancillary dot is increased. These might be
possible with different materials, such as II-VI based systems.

This work is supported in part by Nation Science Foundation Grants
No. PHY0653373 and No. DMR0608501.
\bibliographystyle{apsrev}


\end{document}